\documentclass[english,10pt,a4paper,twocolumn]{article}
\usepackage[T1]{fontenc}
\usepackage[left=2cm, right=2cm, top=4cm, bottom=2cm]{geometry}
\usepackage{graphicx}
\usepackage{mathtools}
\usepackage{amssymb}
\usepackage{amsthm}
\usepackage{babel}
\usepackage{titlesec}
\usepackage{color}
\usepackage{flushend}

\renewcommand{\thesection}{\Roman{section}}        % I, II, III, ...
\renewcommand{\thesubsection}{\Alph{subsection}}   % A, B, C, ...
\renewcommand{\thesubsubsection}{\alph{subsubsection}} % a, b, c

\titleformat{\section}[block]{\centering\bfseries\large}{\thesection.}{1em}{}
\titleformat{\subsection}[block]{\centering\bfseries\normalsize}{\thesubsection.}{1em}{}
\titleformat{\subsubsection}[block]{\centering\bfseries\normalsize}{\thesubsubsection.}{1em}{}

\title{Traversable Wormholes with Non-Exotic Matter: The Role of Higher Curvature Corrections}
\author{M Daniel Ranjan$^1$, Soumya Chakrabarti$^2$, Sanjit Das$^{1*}$\vspace{0.25cm}\\
$^1${\em School of Advanced Sciences, Vellore Insititute of Technology(VIT),}\\{\em Vandalur-Kelambakkam road, Chennai-600127, Tamil Nadu, India}\vspace{0.25cm}\\
$^2${\em School of Advanced Sciences, Vellore Institute of Technology (VIT).}\\
{\em Thiruvalam Road, Katpadi, Vellore, Tamil Nadu, 632014, India\vspace{0.25cm}}\\
$^*$E-mail: {\em sanjit.das@vit.ac.in}}
\date{}
\begin{document}
	\maketitle	

    %Abstract-----------------------------------------------------

    \begin{abstract}
        In this paper, we explore wormhole solutions in a higher–derivative theory of gravity where the action depends not only on the Ricci scalar R, but also on its d’Alembertian, $\Box R$. Such $f(R,\Box R)$ models are motivated by quantum corrections to general relativity and naturally extend the space of possible gravitational geometries. Our goal is to examine whether traversable wormholes can exist in this framework and to understand the role of higher–order curvature terms in supporting them. We derive the field equations for a static, spherically symmetric wormhole and study their solutions using both analytical arguments and numerical methods. Particular attention is given to the classical energy conditions, which are usually violated in wormhole physics. We find that the higher–derivative corrections can effectively contribute to the stress–energy tensor, reducing the amount of exotic matter required at the throat, and in some cases eliminating the need for it altogether.
    \end{abstract}
	
	%Section 1----------------------------------------------------
	
	\section{Introduction}
General Relativity (GR) admits a rich spectrum of compact and exotic spacetime configurations arising from the nonlinear nature of the gravitational field equations. Beyond conventional compact objects such as neutron stars and black holes, the theory allows solutions characterized by horizons, singularities, nontrivial topology, and extreme curvature effects. Examples include black holes with null boundaries, ultra-compact horizonless configurations approaching the Buchdahl limit, and spacetime geometries supported by unconventional stress–energy sources. Among these theoretical possibilities, wormholes occupy a distinctive position, as they represent topologically nontrivial structures connecting separate regions of spacetime through a geometric bridge\cite{visser1995lorentzian}.

Wormholes are spacetime configurations that connect two distant regions of the Universe through a geometric shortcut. The concept first appeared in the work of Einstein and Rosen\cite{01einstein1935particle}, who constructed what is now known as the Einstein-Rosen bridge. Later, Wheeler introduced the term wormhole\cite{02wheeler1957nature}, and since then such geometries have attracted sustained interest in gravitational physics. A major development occurred when Morris and Thorne\cite{03morris1988wormholes} provided the first detailed construction of a static, spherically symmetric traversable wormhole within the framework of GR. Their model established the geometric conditions required for traversability and initiated extensive research into the physical viability of such objects. More recently, such studies have expanded to testing the existence of wormholes as potential black hole mimickers\cite{de2020general,de2021testing}.

A well-known limitation of the Morris–Thorne wormhole in GR is the unavoidable violation of the energy conditions at or near the throat. In particular, the null energy condition (NEC) must be violated in order to satisfy the flare-out condition, implying the existence of so-called \emph{exotic matter}. However, classical forms of matter observed in nature obey the standard energy conditions, and only certain quantum effects are known to permit local violations. The requirement of exotic matter therefore represents a significant obstacle to the physical realization of traversable wormholes. A central objective in modern wormhole research is to minimize, localize, or altogether eliminate the need for exotic matter. Several approaches have been proposed toward this goal, including thin-shell constructions\cite{04visser1989traversable}, quantum corrections\cite{05garattini2019casimir, 06garattini2025rotating}, scalar-field models\cite{08coule1990wormholes,07popov2001vacuum}, higher-dimensional frameworks\cite{10clement1984class}, and various extensions of general relativity\cite{de2021reconstructing,battista2024generalized}. Among these possibilities, modified gravity theories offer a particularly appealing avenue. In such theories, additional geometric contributions appear in the effective gravitational field equations, which can modify the energy-condition balance without requiring exotic matter sources\cite{capozziello2014energy}. In this way, the apparent violation of energy conditions may arise from curvature terms rather than from pathological matter components.

Modified gravity models were originally motivated by cosmological considerations, such as explaining the late-time accelerated expansion of the Universe, addressing early-time singularities, and providing alternatives to dark energy within the standard $\Lambda$CDM paradigm\cite{kerner1982cosmology,buchdahl1970non,capozziello2002curvature,cognola2006dark}. Beyond cosmology, these theories have proven useful in the study of compact objects and nontrivial spacetime topologies. In the context of wormholes, numerous investigations have considered the Morris–Thorne geometry within extended theories such as $f(R)$\cite{lobo2009wormhole}, $f(R,\mathcal{T})$\cite{azizi2013wormhole}, $f(T)$\cite{sharif2014f}, Einstein-Cartan theory\cite{di2017spin}, Einstein–Gauss–Bonnet gravity\cite{ahmed2022wormhole}, and related models, demonstrating that higher-curvature contributions can effectively relax or redistribute the energy-condition requirements.

In the present work, we investigate traversable wormhole geometries in the framework of $f(R,\Box R)$ gravity, where the action depends not only on the Ricci scalar $R$ but also on its d'Alembertian $\Box R$. The inclusion of higher-derivative curvature corrections introduces additional dynamical degrees of freedom that can significantly influence the effective stress–energy content of the spacetime. We show that these higher-curvature corrections can significantly modify the energy-condition requirements for traversable wormholes. For several representative functional forms of the theory, we find that the null energy condition (NEC) can be satisfied in an extended regions of the spacetime when the geometric higher-derivative terms contribute effectively to the gravitational dynamics. Furthermore, we demonstrate that localized radial deformations of the Morris–Thorne geometry and the introduction of explicit time evolution through a scale factor can further improve the NEC behavior, in certain parameter regimes rendering it non-negative throughout the radial domain. These results indicate that the apparent need for exotic matter can, in part, be transferred to purely geometric higher-curvature contributions.  \\

The paper is organized as follows. In Sec.~\ref{sec: bas fr}, we present the basic framework of Morris–Thorne wormholes in $f(R,\Box R)$ gravity and derive the corresponding field equations. In Sec.~\ref{sec: ener cond}, we examine the null energy condition for several representative functional forms of $f(R,\Box R)$. In Sec.~\ref{sec: rad pert}, we introduce a radial deformation of the wormhole geometry and analyze its impact on the energy conditions. In Sec.~\ref{sec: time evol}, we extend the analysis to a time-dependent generalization of the deformed geometry and study how explicit temporal evolution influences the NEC behavior.
	
	%Section 2----------------------------------------------------
	
	\section{Basic Framework of $f(R, \Box R)$ Gravity}\label{sec: bas fr}
	
	We begin by considering the action for \( f(R, \Box R) \) gravity, which extends the Einstein--Hilbert action beyond the standard \( f(R) \) framework by incorporating higher-order curvature corrections through the d'Alembertian of the Ricci scalar \cite{gottlober1990sixth,iihoshi2010mutated,yousaf2018energy,carloni2019cosmology}. Such extensions are well motivated in the context of effective gravitational theories, where higher-derivative terms can arise naturally and provide additional degrees of freedom without the explicit introduction of scalar fields. In particular, the inclusion of the \( \Box R \) term allows the geometric sector itself to account for corrections that may otherwise be attributed to additional fields, and can potentially shift the violation of energy conditions from the matter sector to the effective curvature contributions. However, this increased generality comes at the cost of mathematical complexity, and higher-derivative terms may introduce instabilities depending on the specific functional form of the model.

    The action can be written as
    \begin{equation}
	       S = \int d^4x\, \sqrt{-g} \left[f(R, \Box R) + \mathcal{L}_m\right],
	       \label{eqn:action}
    \end{equation}
    where \( g \) denotes the determinant of the metric tensor \( g_{\mu\nu} \), \( f(R, \Box R) \) is an arbitrary function of the Ricci scalar \( R \) and its d'Alembertian \( \Box R = g^{\mu\nu}\nabla_\mu\nabla_\nu R \), and \( \mathcal{L}_m \) represents the matter Lagrangian density.

    Variation of the action with respect to the metric \( g_{\mu\nu} \) leads to the modified field equations
    \begin{eqnarray}
	       \mathbb{G} G_{\mu\nu} &=& \tfrac{1}{2} g_{\mu\nu}(f - \mathbb{G} R)
	       + \mathbb{G}_{;\mu\nu} - g_{\mu\nu}\Box \mathbb{G} \nonumber \\
	       && - \tfrac{1}{2} g_{\mu\nu}\!\left(\mathcal{F}_{;\gamma} R^{;\gamma} + \mathcal{F}\Box R\right) \nonumber \\
	       && + \mathcal{F}_{;(\mu} R_{;\nu)} + T_{\mu\nu},
	       \label{eqn:field1}
    \end{eqnarray}
    where \( G_{\mu\nu} \) is the Einstein tensor and \( T_{\mu\nu} \) is the energy--momentum tensor of matter,
    \[
        T^\mu_{\;\nu} = \mathrm{diag}[-\rho,\, p_r,\, p_t,\, p_t],
    \]
    and the quantities
    \begin{equation*}
	       \mathbb{G} = \frac{\partial f}{\partial R} + \Box \mathcal{F}, 
	       \qquad
	       \mathcal{F} = \frac{\partial f}{\partial (\Box R)}
    \end{equation*}
    have been introduced for convenience.

    To investigate wormhole geometries within this framework, we consider a static and spherically symmetric spacetime described by the Morris--Thorne metric,
    \begin{equation}
	       ds^2 = -e^{\Phi(r)} dt^2 + \frac{dr^2}{1 - \tfrac{b(r)}{r}} + r^2 (d\theta^2 + \sin^2\theta\, d\phi^2),
            \label{eqn:mtmetric}
    \end{equation}
    where \( \Phi(r) \) and \( b(r) \) denote the redshift and shape functions, respectively. For simplicity, and to ensure a zero-tidal-force configuration, we assume the redshift function to be constant, setting
    \[
        \Phi(r) = 0.
    \]
    This choice eliminates gravitational redshift effects and significantly simplifies the analysis, allowing the field equations to be expressed solely in terms of the shape function \( b(r) \) and the derivatives of \( f(R, \Box R) \).

	%Section 3----------------------------------------------------
	
	\section{Energy Conditions}\label{sec: ener cond}
		
	Energy conditions arise from the Raychaudhuri equations, which describe how nearby paths of light or matter move through spacetime. They provide general guidelines for what kinds of matter and energy distributions are physically reasonable, and how these influence the geometry of spacetime. In simple terms, the energy conditions tell us whether gravity behaves in a “normal” attractive way or not.
	
	There are four commonly used energy conditions. The null energy condition (NEC), the weak energy condition (WEC), the strong energy condition (SEC), and the dominant energy condition (DEC).
	In this paper, we focus on the Null Energy Condition, given by

    \begin{equation*}
        \rho + p_i \geq 0
    \end{equation*}
    
    since it plays a key role in understanding wormhole geometries. Morris and Thorne showed that, in Einstein’s theory of gravity, a stable and traversable wormhole must violate this condition\cite{03morris1988wormholes}. This means that the matter required to hold the wormhole open must behave in ways that ordinary matter cannot—it is called exotic matter.
	
	However, when we move beyond Einstein’s theory and consider modified gravity models, the situation changes. In these theories, additional geometric terms appear in the gravitational field equations, and these terms can effectively mimic violations of the energy conditions. In other words, the apparent violation may come from the curvature of spacetime itself, not from any exotic form of matter. This makes it possible to construct wormhole solutions that are supported by ordinary matter, with the modified gravity providing the necessary geometric contribution.
	
	In the following, we will study how the energy conditions behave for matter in three different models of $ f(R, \Box R) $ gravity. This will help us understand whether the violation of the NEC can be attributed to the matter or to the geometry of spacetime.
	
	\subsection{Model I}
	
	The first $f(R,\Box R)$ model considered in this article includes a quadratic curvature correction together with the higher-order derivative term\cite{yousaf2018energy} and is given by 
    \begin{equation}
        f(R,\Box R) = R + k_{1}R^{2} + k_{2} R \Box R ,
        \label{eqn:model I}
    \end{equation}
    where $k_{1}$ and $k_{2}$ are dimensionful real constants.

    \begin{figure}[h!]
	       \centering
	       \includegraphics[width=0.9\linewidth]{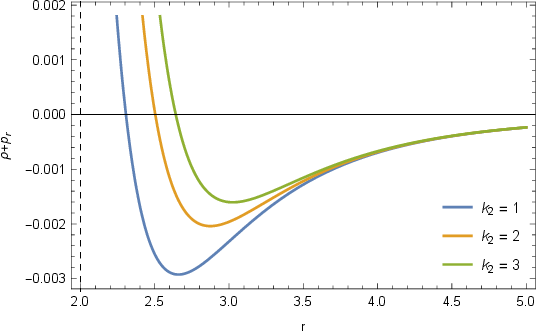}
	       \includegraphics[width=0.9\linewidth]{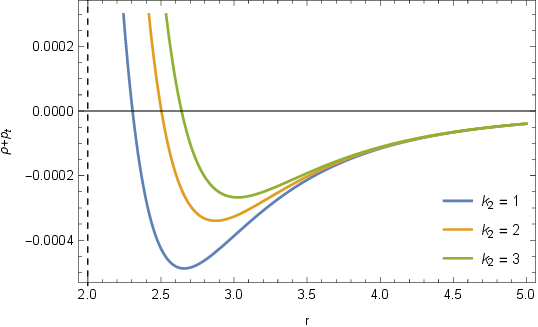}
	       \caption{\textbf{Model I:} Plots of radial $(\rho+p_r)$ and tangential $(\rho+p_t)$ null energy condition profiles for the parameters $b_{0}=2$ and $k_{1}=1$.}
	       \label{fig:Ml1NEC1}
    \end{figure}

    Figure~\ref{fig:Ml1NEC1} illustrates the behavior of the null energy condition for Model I. We choose the shape function profile $b(r)=\frac{{b_0}^2}{r}$ throughout this paper. The NEC profiles are plotted for different values of the parameter $k_{2}$ while keeping $b_{0}=2$ and $k_{1}=1$. It is observed that both quantities become negative in the vicinity of the wormhole throat, indicating a violation of the NEC in that region. As $k_{2}$ increases, the minimum values of $(\rho+p_r)$ and $(\rho+p_t)$ increase, and the magnitude of the violation decreases. At larger radial distances, both quantities gradually approach zero, suggesting that the violation weakens away from the throat.
	
	\subsection{Model II}
	
	In the second $f(R,\Box R)$ model, the quadratic curvature term is extended to include a cubic contribution while retaining the higher-derivative $R\Box R$ correction\cite{yousaf2018energy}. The model is given by
    \begin{equation}
        f(R,\Box R)=R + k_{1} R^{2}\left(1 + k_{3} R\right) + k_{2} R \Box R ,
        \label{eqn:model II}
    \end{equation}
    where $k_{1}$, $k_{2}$, and $k_{3}$ are real dimensionful constants.

    \begin{figure}[h!]
	       \centering
	       \includegraphics[width=0.9\linewidth]{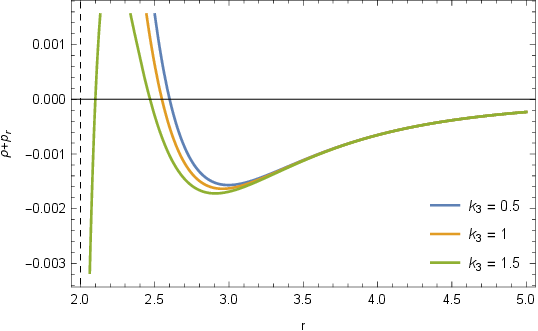}
	       \includegraphics[width=0.9\linewidth]{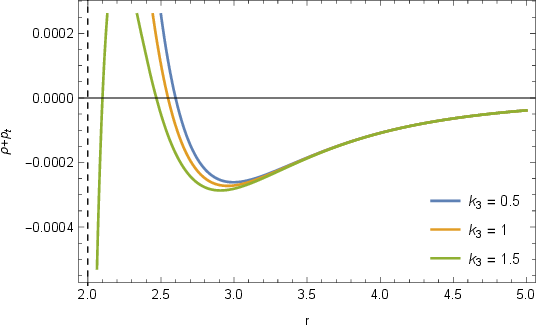}
	       \caption{\textbf{Model II:} Plots of radial $(\rho+p_r)$ and tangential $(\rho+p_t)$ null energy condition profiles for the parameters $b_{0}=2$, $k_{1}=3$, and $k_{2}=1$.}
	       \label{fig:Ml2NEC2}
    \end{figure}

    \begin{figure}[h!]
	       \centering
	       \includegraphics[width=0.9\linewidth]{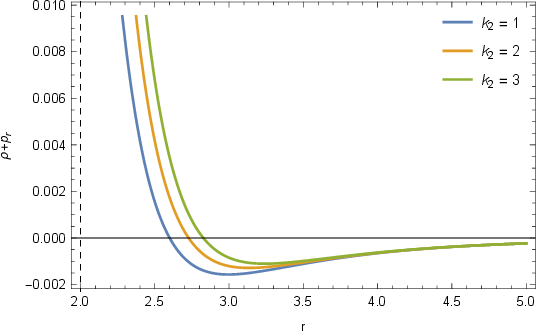}
	       \includegraphics[width=0.9\linewidth]{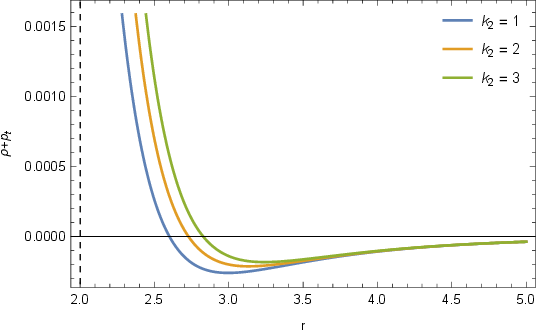}
	       \caption{\textbf{Model II:} Plots of radial $(\rho+p_r)$ and tangential $(\rho+p_t)$ null energy condition profiles for the parameters $b_{0}=2$, $k_{1}=3$, and $k_{3}=1$.}
	       \label{fig:Ml2NEC3}
    \end{figure}

    Figures~\ref{fig:Ml2NEC2} and \ref{fig:Ml2NEC3} illustrate the behavior of the radial and tangential null energy conditions for Model~II. The upper panels correspond to the radial NEC $(\rho+p_r)$, while the lower panels represent the tangential NEC $(\rho+p_t)$.

    Figure~\ref{fig:Ml2NEC2} shows the NEC profiles for fixed $k_{2}=1$ and varying values of $k_{3}$. It is observed that increasing $k_{3}$ leads to a deeper negative minimum of both $\rho+p_r$ and $\rho+p_t$, thereby reducing the region in which the NEC is satisfied.

    Figure~\ref{fig:Ml2NEC3} presents the corresponding profiles for $k_{3}=0.5$ while varying the higher-derivative contribution. In contrast to the behavior induced by $k_{3}$, increasing $k_{2}$ shifts the curves towards larger values of $\rho+p_r$ and $\rho+p_t$, reducing the magnitude of the NEC violation and enlarging the region where the NEC is satisfied. At sufficiently large radial distances, both quantities gradually approach zero.
	
	\subsection{Model III}
	
	The third $f(R,\Box R)$ model introduces a non-polynomial modification to the previous models through an exponential dependence on $R$, while retaining the higher-derivative $R\Box R$ contribution\cite{yousaf2018energy}. The model is defined as
    \begin{eqnarray}
        f(R, \Box R) &=& R + k_{1} R \left( \exp\!\left[-\frac{R}{k_{3}}\right] - 1 \right) \nonumber \\
        && \qquad\qquad + k_{2} R \Box R ,
        \label{eqn:model III}
    \end{eqnarray}
    where $k_{1}$, $k_{2}$, and $k_{3}$ are real constants.

    \begin{figure}[h!]
	       \centering
	       \includegraphics[width=0.9\linewidth]{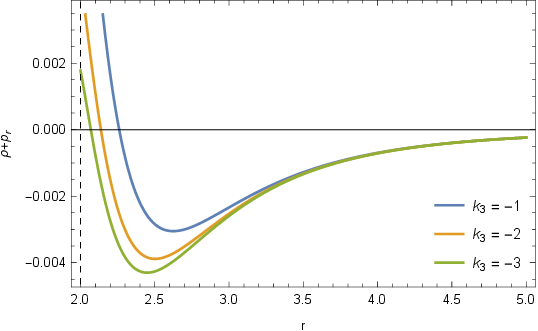}
	       \includegraphics[width=0.9\linewidth]{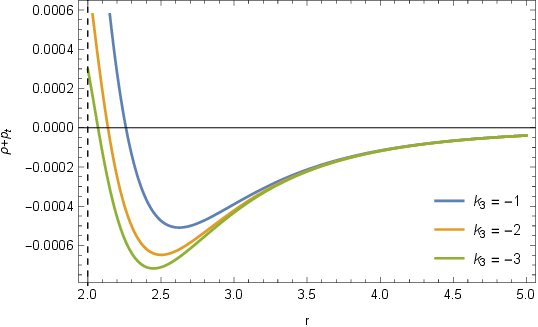}
	       \caption{\textbf{Model III:} Plots of radial $(\rho+p_r)$ and tangential $(\rho+p_t)$ null energy condition profiles for the parameters $b_{0}=2$, $k_{1}=1$, and $k_{2}=1$.}
	       \label{fig:Ml3NEC1}
    \end{figure}

    \begin{figure}[h!]
	       \centering
	       \includegraphics[width=0.9\linewidth]{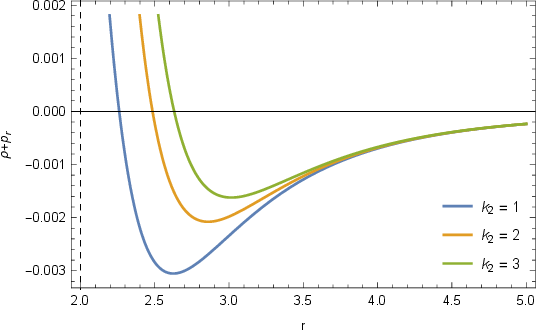}
	       \includegraphics[width=0.9\linewidth]{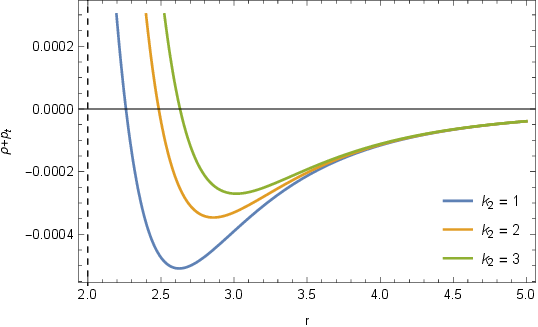}
	       \caption{\textbf{Model III:} Plots of radial $(\rho+p_r)$ and tangential $(\rho+p_t)$ null energy condition profiles for the parameters
           $b_{0}=2$, $k_{1}=1$, and $k_{3}=-1$.}
	       \label{fig:Ml3NEC2}
    \end{figure}

    \begin{figure*}[h!]
		\centering
		\includegraphics[width=0.45\linewidth]{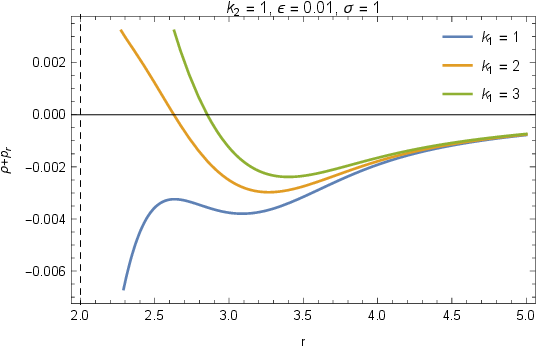}
		\includegraphics[width=0.45\linewidth]{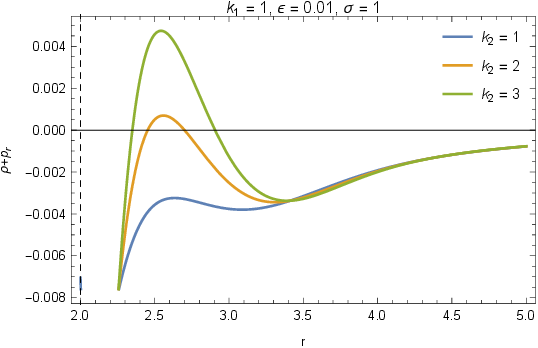}
		\includegraphics[width=0.45\linewidth]{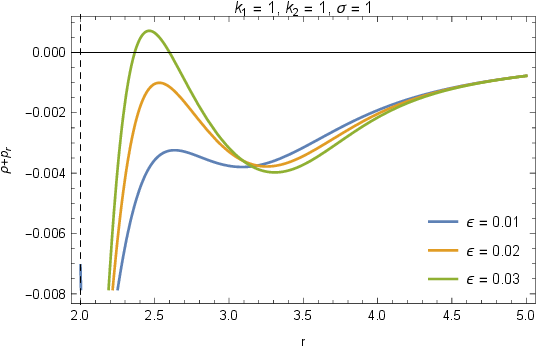}
		\includegraphics[width=0.45\linewidth]{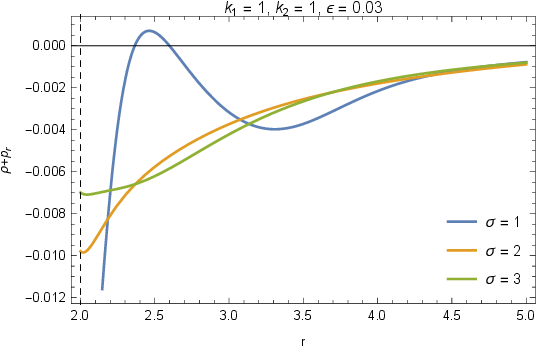}
		\caption{Plots of Radial Null Energy condition for model I with deformation. Here $b_{0} = 2$.}
		\label{fig:PMl1NEC1}
	\end{figure*}
	
	\begin{figure*}[h!]
		\centering
		\includegraphics[width=0.45\linewidth]{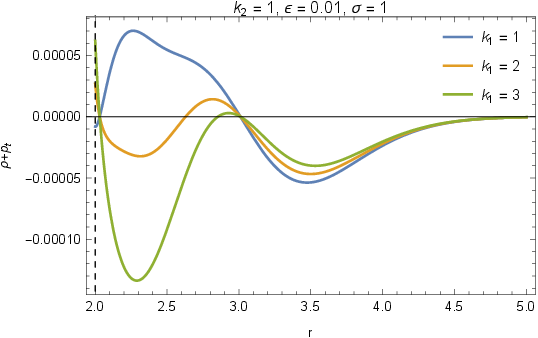}
		\includegraphics[width=0.45\linewidth]{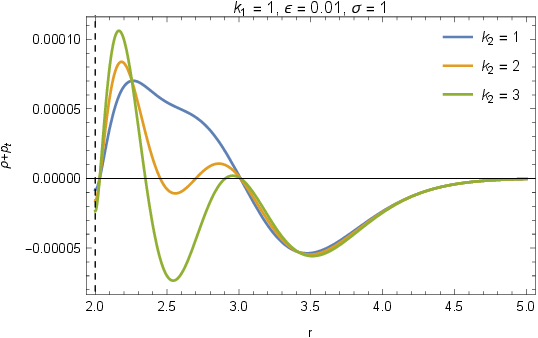}
		\includegraphics[width=0.45\linewidth]{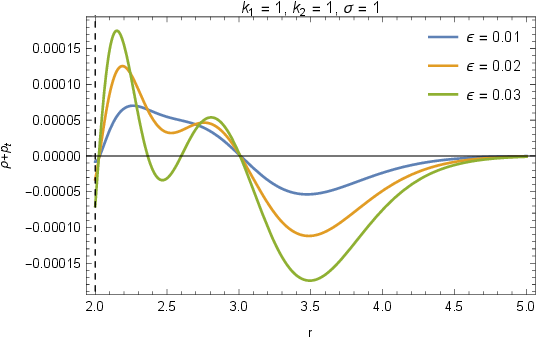}
		\includegraphics[width=0.45\linewidth]{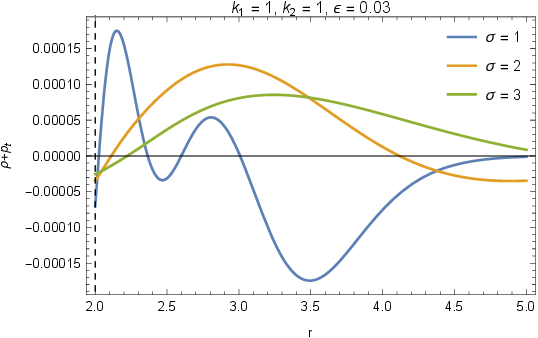}
		\caption{Plots of Tangential Null Energy condition for model I with deformation. Here $b_{0} = 2$.}
		\label{fig:PMl1NEC2}
	\end{figure*}

    Figures~\ref{fig:Ml3NEC1} and \ref{fig:Ml3NEC2} illustrate the behavior of the radial and tangential null energy conditions for Model~III. The upper panels correspond to the radial NEC $(\rho+p_r)$, while the lower panels represent the tangential NEC $(\rho+p_t)$.

    Figure~\ref{fig:Ml3NEC1} shows the NEC profiles for fixed $k_{2}=1$ and varying values of $k_{3}$. It is observed that increasing the magnitude of the exponential curvature contribution leads to a deeper negative minimum of both $\rho+p_r$ and $\rho+p_t$, thereby reducing the radial interval over which the NEC is satisfied.

    Figure~\ref{fig:Ml3NEC2} presents the corresponding behavior for fixed $k_{3}=-1$ and varying values of the higher-derivative parameter $k_{2}$. In this case, increasing $k_{2}$ shifts the curves towards larger values of $\rho+p_r$ and $\rho+p_t$, reducing the magnitude of the NEC violation and enlarging the region where the NEC is satisfied. At larger radial distances, both quantities gradually approach zero.
	
	%Section 4----------------------------------------------------
	
	\section{Morris-Thorne Wormhole with Radial Deformation}\label{sec: rad pert}
	
\begin{figure}[h!]
        \centering
        \includegraphics[width=0.9\linewidth]{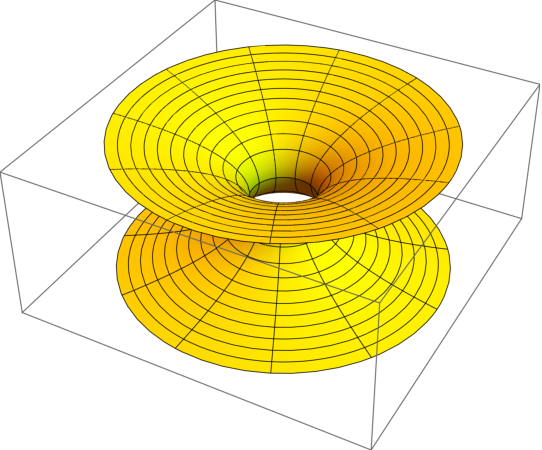}
        \caption{Embedding diagram of the wormhole metric with radial deformation}
        \label{fig:ped}
    \end{figure}

	As discussed in the previous section, although the NEC is locally satisfied near the throat, the region in which this occurs is too narrow to be physically significant. To enlarge this region, we modify the wormhole metric Eqn. (\ref{eqn:mtmetric}) by replacing every occurrence of \( r \) with \( r + \epsilon\, g(r) \). Although this may appear as if the radial coordinate is being perturbed by an arbitrary function, no coordinate transformation is involved; instead, the substitution directly alters the metric itself. The revised line element is
	\begin{equation}
		ds^{2} = -e^{\Phi(r)} dt^{2}
		+ \frac{dr^{2}}{1 - \frac{b(r)}{\tilde{r}}}
		+ \tilde{r}^{2} d\theta^{2}
		+ \tilde{r}^{2} \sin^{2}\theta\, d\phi^{2},
		\label{eqn:pertmetric}
	\end{equation}
	where \( \tilde{r} = r + \epsilon\, g(r) \), \( \epsilon \) is a real parameter with dimension of length, and \( g(r) \) is a random function of $r$. In this work, we choose
	\begin{equation}
		g(r) = \exp\!\left[-\left( \frac{r - b_{0}}{\sigma} \right)^{2} \right],
	\end{equation}
	where \( \sigma \) is a real parameter with length dimension. This choice produces a localized Gaussian deformation near the throat, with its width controlled by the parameter \( \sigma \). The corresponding embedding diagram for the modified metric is shown in Figure~\ref{fig:ped}

	\subsection{Model I}
	
	Figures~\ref{fig:PMl1NEC1} and \ref{fig:PMl1NEC2} present the profiles of the radial and tangential null energy conditions for the wormhole in \( f(R,\Box R) \) gravity Model~I given by Eqn. (\ref{eqn:model I}) under radial deformation. The plots indicate the presence of a localized region in which the NEC is satisfied, forming a pocket where the NEC holds and thereby reducing the amount of exotic matter required to sustain the wormhole geometry. The extent and location of this region depend on the parameters \( k_{1} \), \( k_{2} \), \( \epsilon \), and \( \sigma \). Variations in \( k_{1} \) and \( k_{2} \) primarily modify the amplitude of the NEC profiles and slightly shift the radial position where the curves cross the zero line. The parameter \( \epsilon \) mainly affects the magnitude of the NEC functions, producing only a mild change in the size of the region where the NEC is satisfied. In contrast, the width parameter \( \sigma \) has a more pronounced influence: increasing \( \sigma \) broadens the region where the tangential NEC is satisfied, while the radial NEC tends to remain negative over a slightly wider interval near the throat. Consequently, the deformation parameters control both the magnitude and the radial distribution of NEC violation in the deformed configuration.
	
	\subsection{Model II}
	
	\begin{figure*}[h!]
		\centering
		\includegraphics[width=0.45\linewidth]{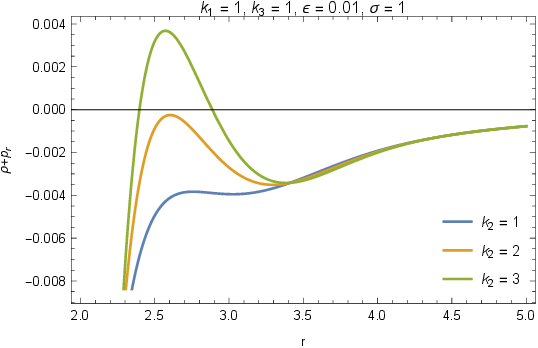}
		\includegraphics[width=0.45\linewidth]{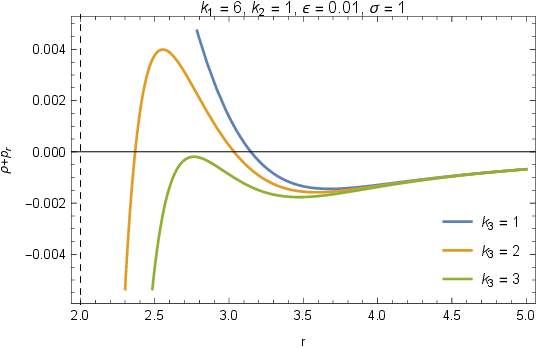}
		\includegraphics[width=0.45\linewidth]{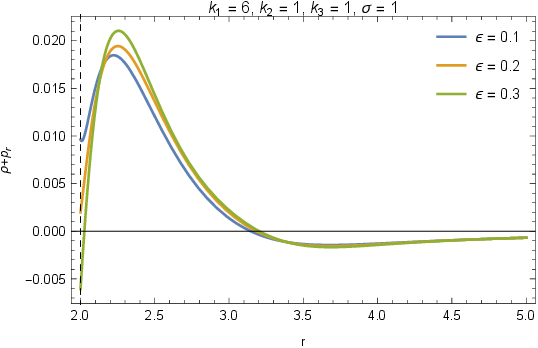}
		\includegraphics[width=0.45\linewidth]{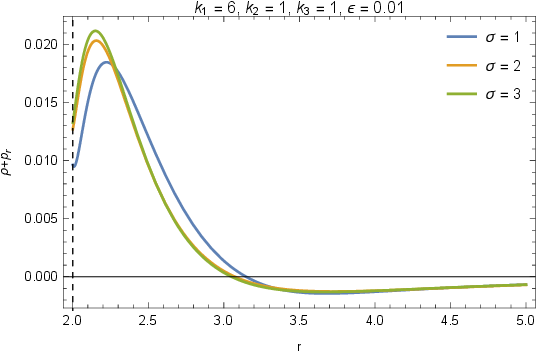}
		\caption{Plots of Radial Null Energy condition for model II with deformation. Here $b_{0} = 2$.}
		\label{fig:PMl2NEC1}
	\end{figure*}
	
	\begin{figure*}[h!]
		\centering
		\includegraphics[width=0.45\linewidth]{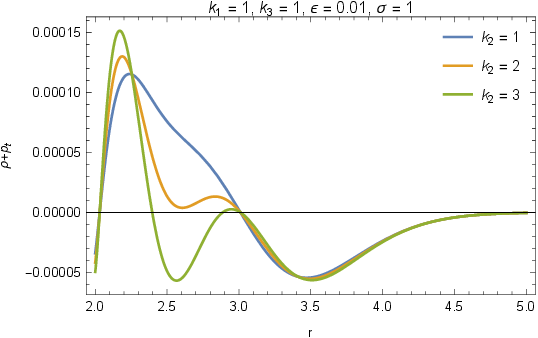}
		\includegraphics[width=0.45\linewidth]{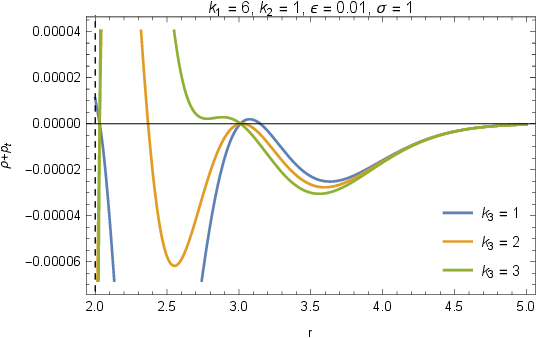}
		\includegraphics[width=0.45\linewidth]{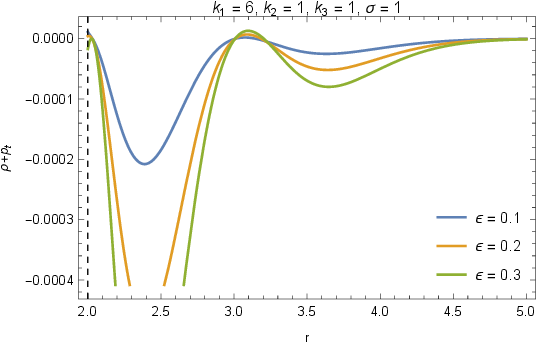}
		\includegraphics[width=0.45\linewidth]{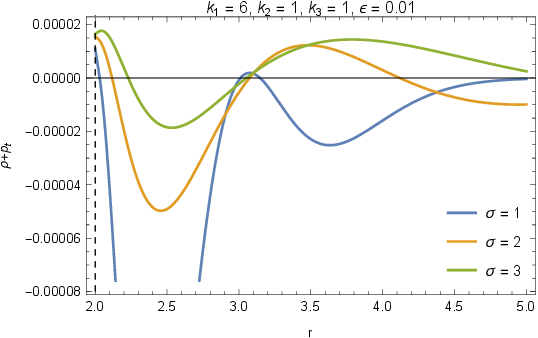}
		\caption{Plots of Tangential Null Energy condition for model II with deformation. Here $b_{0} = 2$.}
		\label{fig:PMl2NEC7}
	\end{figure*}
	
    Figures~\ref{fig:PMl2NEC1} and \ref{fig:PMl2NEC7} illustrate the profiles of the radial and tangential null energy conditions for the wormhole in \( f(R,\Box R) \) gravity Model~II given by Eqn. (\ref{eqn:model II}) under radial deformation. The plots demonstrate that the inclusion of the cubic curvature term modifies the distribution of NEC violation in the wormhole spacetime. In particular, the contribution from the cubic curvature parameter \(k_{3}\) tends to reduce the radial extent of the region where the radial NEC is satisfied, while simultaneously enlarging the region where the tangential NEC holds. Variations in the parameters \(k_{2}\) and \(k_{3}\) primarily affect the extrema of the NEC profiles. Changes to the parameter \( \epsilon \) while modifying the magnitude of the NEC functions does not significantly alter the overall structure of the curves. Furthermore, the width parameter \( \sigma \) exhibits different effects on the two components of the NEC: the radial NEC remains largely insensitive to variations in \( \sigma \), whereas increasing \( \sigma \) broadens the region where the tangential NEC is satisfied. These results indicate that the cubic curvature correction plays an important role in determining the radial distribution of NEC violation in the deformed Model~II wormhole configuration.
	
	\subsection{Model III}
	
\begin{figure*}
		\centering
		\includegraphics[width=0.45\linewidth]{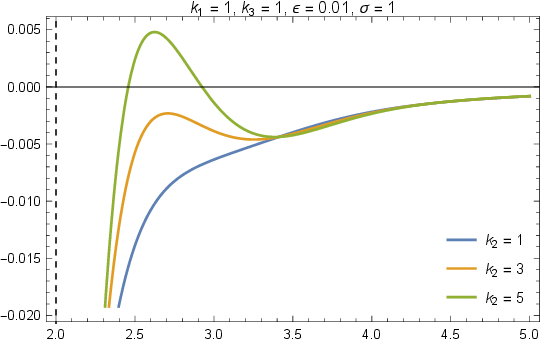}
		\includegraphics[width=0.45\linewidth]{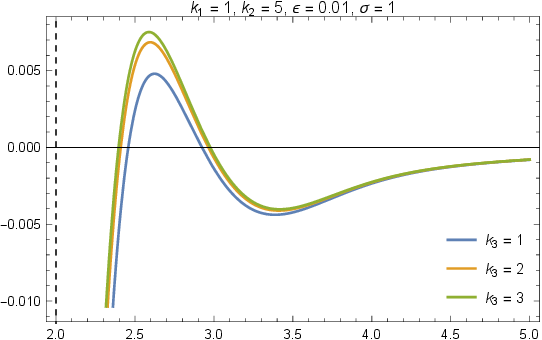}
		\includegraphics[width=0.45\linewidth]{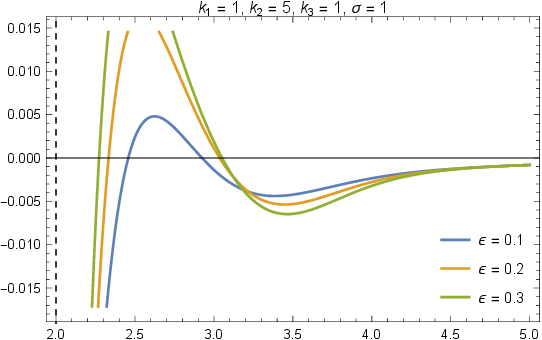}
		\includegraphics[width=0.45\linewidth]{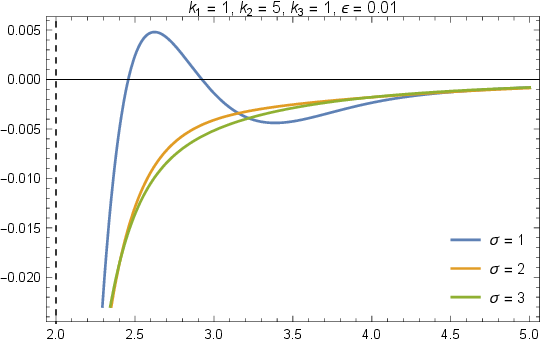}
		\caption{Plots of Radial Null Energy condition for model III with deformation. Here $b_{0} = 2$.}
		\label{fig:PMl3NEC1}
	\end{figure*}

\begin{figure*}
		\centering
		\includegraphics[width=0.45\linewidth]{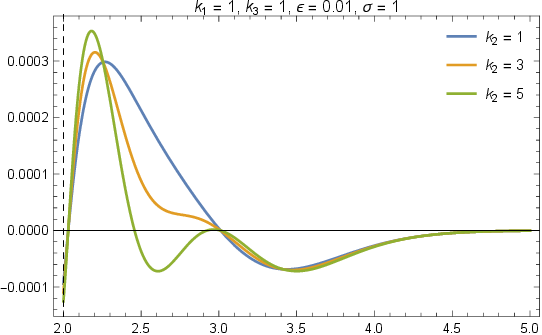}
		\includegraphics[width=0.45\linewidth]{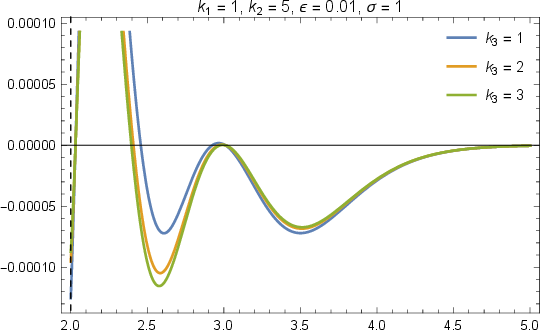}
		\includegraphics[width=0.45\linewidth]{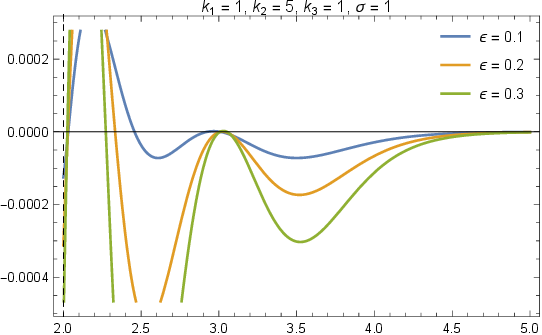}
		\includegraphics[width=0.45\linewidth]{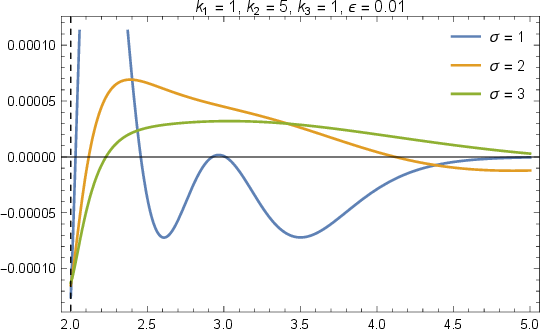}
		\caption{Plots of Tangential Null Energy condition for model III with deformation. Here $b_{0} = 2$.}
		\label{fig:PMl3NEC2}
	\end{figure*}

	Figures~\ref{fig:PMl3NEC1} and \ref{fig:PMl3NEC2} illustrate the profiles of the radial and tangential null energy conditions for the wormhole in \( f(R,\Box R) \) gravity Model~III given by Eqn. (\ref{eqn:model III}) under radial deformation. The plots indicate that an increase in the parameters \(k_{2}\), \(k_{3}\), and \(\epsilon\) increases the radial extent of the region where the radial NEC is satisfied. In contrast, increasing the width parameter \( \sigma \) reduces the size of this region. The behavior of the tangential NEC exhibits an opposite trend: the region where the tangential NEC is satisfied decreases with increasing values of \(k_{2}\), \(k_{3}\), and \(\epsilon\), while larger values of \( \sigma \) enlarge the corresponding region. Thus, among the parameters considered, increasing \( \sigma \) provides the most effective way to broaden the region where the tangential NEC is satisfied in Model~III.
	
    %Section 5----------------------------------------------------
	
	\section{Evolving Morris--Thorne Wormhole with Radial Deformation}\label{sec: time evol}
	
	We now extend the deformed Morris-Thorne wormhole geometry, Eqn. (\ref{eqn:pertmetric}) by introducing an explicit time dependence through a scale factor multiplying the spatial sector of the metric. The resulting line element is given by
	\begin{equation}
		\begin{aligned}
			ds^{2} ={}& - e^{\Phi(r)} \, dt^{2} + a(t)^{2} \left(\right.
			\frac{dr^{2}}{1 - \dfrac{b(r)}{\tilde{r}}} \\
			&\left.+ \tilde{r}^{2} d\theta^{2}
			+ \tilde{r}^{2} \sin^{2}\theta \, d\phi^{2}
			\right),
		\end{aligned}
        \label{eqn: evolving wormhole}
	\end{equation}
	where the function \( a(t) \) governs the explicit time evolution of the spatial geometry. This construction allows us to examine how temporal dynamics influence the energy-condition requirements of the deformed wormhole spacetime.
	
	We adopt the following power-law form for the scale factor:
	\begin{equation}
		a(t) = \omega \, t^{m},
	\end{equation}
	where \( \omega \) is a constant normalization parameter and \( m \) controls the strength of the time dependence. The values \( m = \tfrac{1}{2} \) and \( m = \tfrac{2}{3} \) are considered as representative examples, corresponding to radiation-like and matter-like power-law evolutions, respectively. These choices are employed phenomenologically to explore the sensitivity of the energy conditions to different temporal behaviors of the wormhole geometry.
	
	\subsection{Model I}
	
	\begin{figure}[h!]
		\centering
		\includegraphics[width=0.9\linewidth]{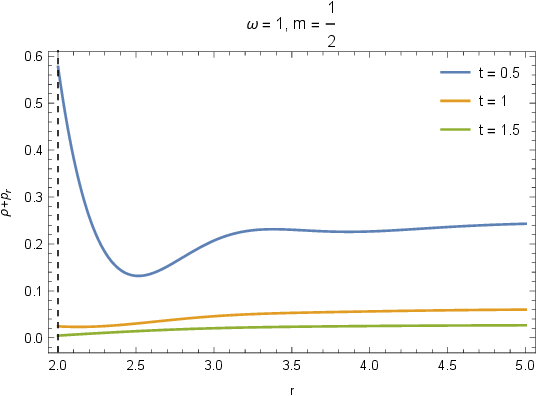}
		\includegraphics[width=0.9\linewidth]{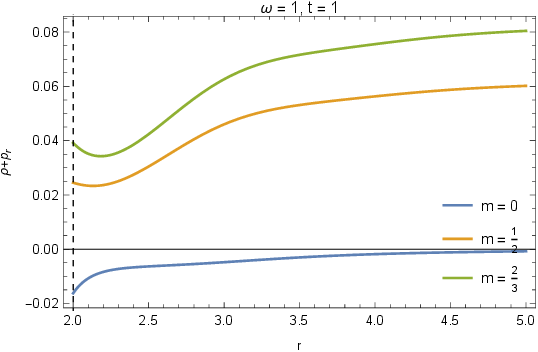}
		\includegraphics[width=0.9\linewidth]{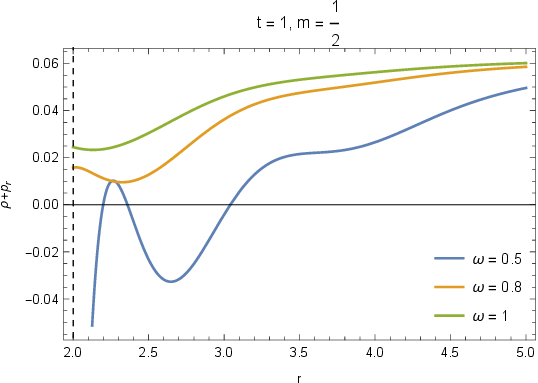}
		\caption{Plots of Radial Null Energy condition for model I with deformation and time evolution. Here $b_{0} = 2$, $k_{1} = 1$, $k_{2} = 1$, $\epsilon = 0.01$ and $\sigma = 1$.}
		\label{fig:EPMl1NEC1}
	\end{figure}
	
	\begin{figure}[h!]
		\centering
		\includegraphics[width=0.9\linewidth]{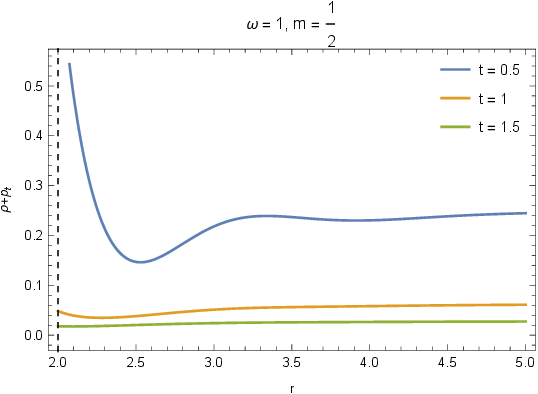}
		\includegraphics[width=0.9\linewidth]{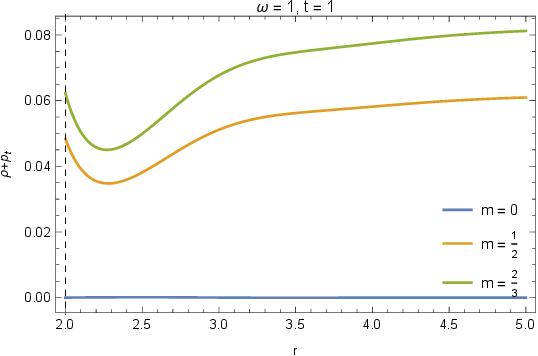}
		\includegraphics[width=0.9\linewidth]{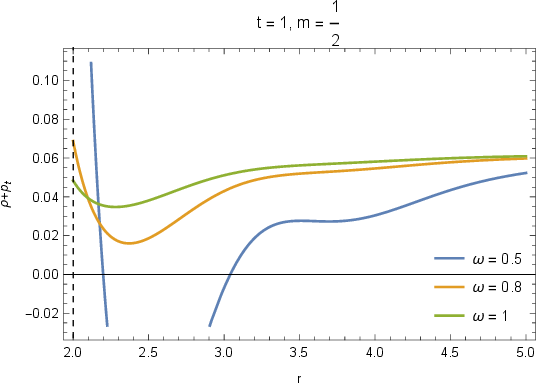}
		\caption{Plots of Tangential Null Energy condition for model I with deformation and time evolution. Here $b_{0} = 2$, $k_{1} = 1$, $k_{2} = 1$, $\epsilon = 0.01$ and $\sigma = 1$.}
		\label{fig:EPMl1NEC2}
	\end{figure}
	
	Figures~\ref{fig:EPMl1NEC1} and \ref{fig:EPMl1NEC2} illustrate the radial and tangential null energy conditions (NEC) for Model~I in the presence of radial deformation and explicit time evolution, for various choices of \( t \), \( m \), and \( \omega \).
	
	The time-evolution plots indicate that at early times the NEC is satisfied over most of the radial domain. As time increases, the NEC near the throat decreases and eventually becomes negative, signaling the emergence of localized NEC violation in the vicinity of the throat. Although the onset of violation occurs close to the throat, the radial extent of the violating region increases gradually with time.
	
	The role of the parameter \( m \) further clarifies this behavior. For \( m = 0 \), the scale factor reduces to a constant \( a(t) = \omega \), and for \( \omega = 1 \) the geometry corresponds to the static deformed wormhole described by Eq.~(\ref{eqn:pertmetric}). In this static limit, the radial NEC is violated throughout the radial range considered, while the tangential NEC exhibits violation only in a restricted region. As \( m \) increases, the magnitude of violation in both the radial and tangential NEC components is progressively reduced, indicating that stronger time dependence improves the overall energy-condition behavior of the spacetime.
	
	A similar trend is observed when varying the normalization parameter \( \omega \) at fixed \( t \) and \( m \). Larger values of \( \omega \), which enhance the contribution of the time-dependent scale factor, lead to an increase in both radial and tangential NEC components. As a result, the severity and radial extent of NEC violation near the throat are further suppressed. These results highlight the mitigating role of time evolution in reducing the exotic matter content required to support the deformed wormhole geometry in Model~I.
	
	\subsection{Model II}
	
	\begin{figure}[h!]
		\centering
		\includegraphics[width=0.9\linewidth]{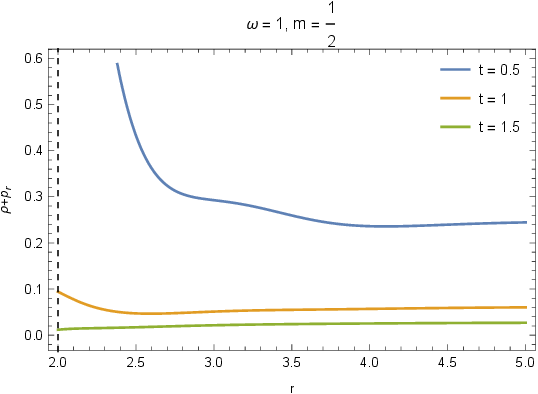}
		\includegraphics[width=0.9\linewidth]{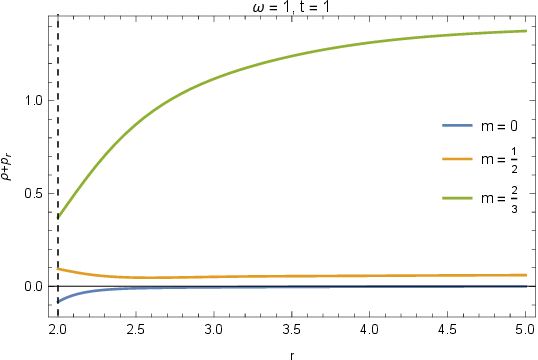}
		\includegraphics[width=0.9\linewidth]{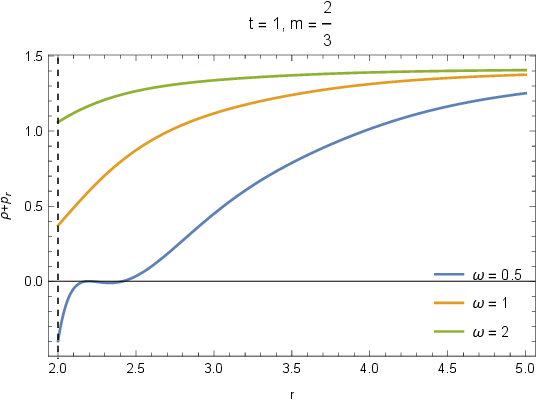}
		\caption{Plots of Radial Null Energy condition for model II with deformation and time evolution. Here $b_{0} = 2$, $k_{1} = 1$, $k_{2} = 1$, $k_{3} = 3$, $\epsilon = 0.01$ and $\sigma = 1$.}
		\label{fig:EPMl2NEC1}
	\end{figure}
	
	\begin{figure}[h!]
		\centering
		\includegraphics[width=0.9\linewidth]{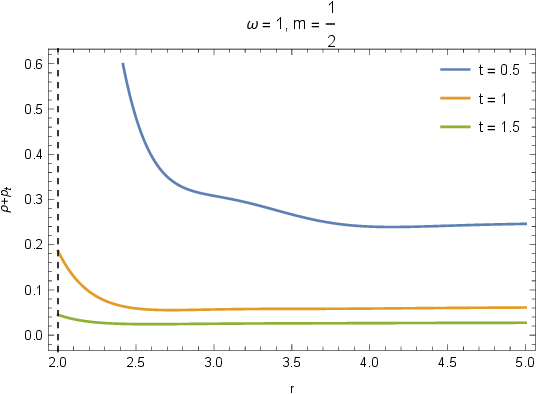}
		\includegraphics[width=0.9\linewidth]{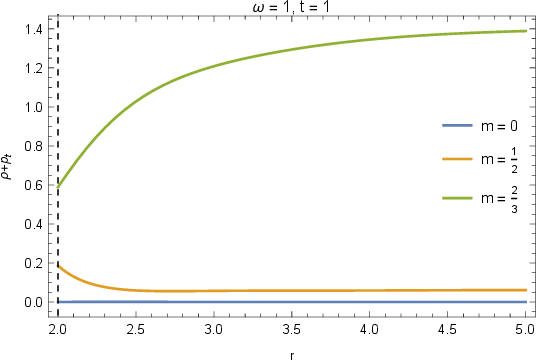}
		\includegraphics[width=0.9\linewidth]{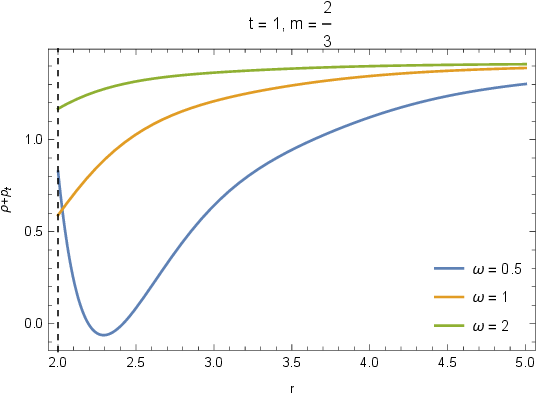}
		\caption{Plots of Tangential Null Energy condition for model II with deformation and time evolution. Here $b_{0} = 2$, $k_{1} = 1$, $k_{2} = 1$, $k_{3} = 3$, $\epsilon = 0.01$ and $\sigma = 1$.}
		\label{fig:EPMl2NEC2}
	\end{figure}
	
	Figures~\ref{fig:EPMl2NEC1} and \ref{fig:EPMl2NEC2} present the radial and tangential null energy conditions for Model~II, including the effects of the deformation and time evolution. The qualitative behavior of the NEC in Model~II closely resembles that observed for Model~I.
	
	At early times, the NEC remains non-negative over a substantial portion of the radial domain. With increasing time, both the radial and tangential NEC decrease near the throat, leading to localized violation. As time evolves further, the region of violation expands outward from the throat, while remaining confined to a comparatively narrow radial range.
	
	The dependence on the parameters \( m \) and \( \omega \) follows the same systematic pattern as in Model~I. The case \( m = 0 \) corresponds to the static deformed configuration, for which the radial NEC is more strongly violated than the tangential component. Increasing \( m \) progressively suppresses the violation in both components. Similarly, larger values of \( \omega \) enhance the NEC contributions, thereby reducing both the magnitude and spatial extent of NEC violation near the throat.
	
	Overall, Model~II exhibits the same physical trends as Model~I, with quantitative differences arising from the additional curvature terms characterizing this model.
	
	\subsection{Model III}
	
	\begin{figure}[h!]
		\centering
		\includegraphics[width=0.9\linewidth]{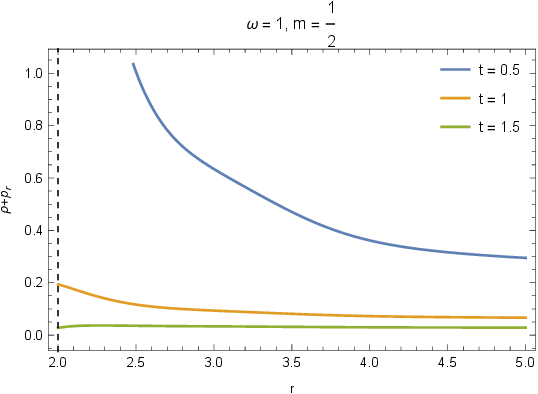}
		\includegraphics[width=0.9\linewidth]{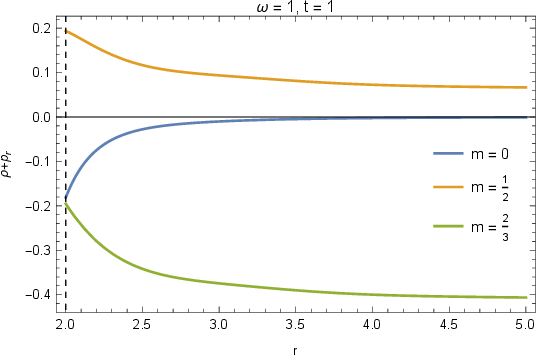}
		\includegraphics[width=0.9\linewidth]{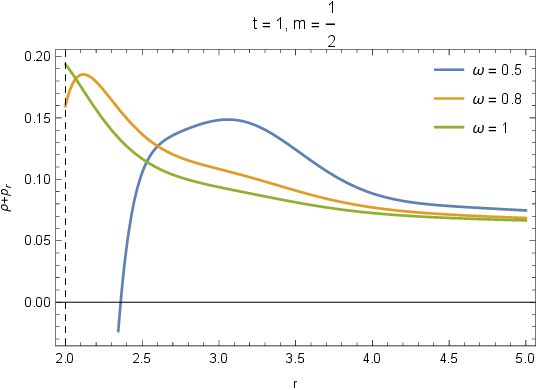}
		\caption{Plots of Radial Null Energy condition for model III with deformation and time evolution. Here $b_{0} = 2$, $k_{1} = 3$, $k_{2} = 1$, $k_{3} = 1$, $\epsilon = 0.01$ and $\sigma = 1$.}
		\label{fig:EPMl3NEC1}
	\end{figure}
	
	\begin{figure}[h!]
		\centering
		\includegraphics[width=0.9\linewidth]{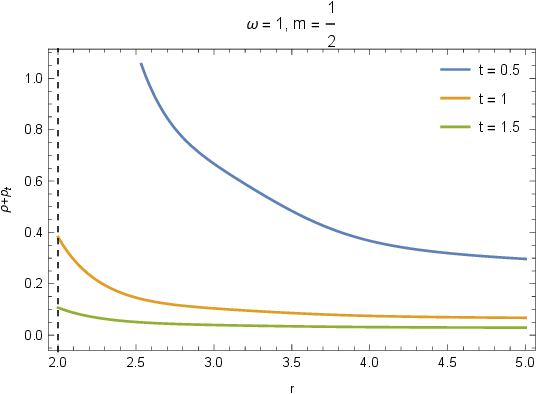}
		\includegraphics[width=0.9\linewidth]{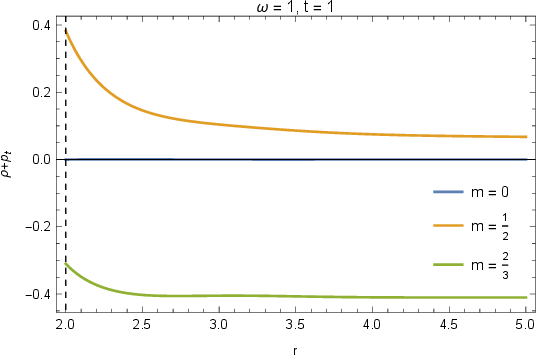}
		\includegraphics[width=0.9\linewidth]{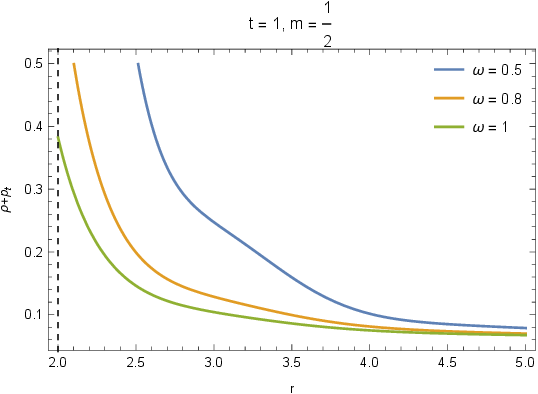}
		\caption{Plots of Tangential Null Energy condition for model III with deformation and time evolution. Here $b_{0} = 2$, $k_{1} = 3$, $k_{2} = 1$, $k_{3} = 1$, $\epsilon = 0.01$ and $\sigma = 1$.}
		\label{fig:EPMl3NEC2}
	\end{figure}
	
	Figures~\ref{fig:EPMl3NEC1} and \ref{fig:EPMl3NEC2} show the radial and tangential null energy conditions for Model~III for different values of \( t \), \( m \), and \( \omega \). Similar to Models~I and II, the time-evolution plots reveal a decrease of the NEC throughout the radial domain as time increases. The onset of NEC violation again occurs first in the vicinity of the throat, and with further increase in time, the region of violation progressively extends to larger radius.
	
	The dependence on the parameter \( m \) exhibits a non-monotonic behavior. The NEC does not attain its maximum value at either the smallest value \( m = 0 \) or the largest value shown \( m = \tfrac{2}{3} \), but instead peaks at the intermediate value \( m = \tfrac{1}{2} \). This indicates that the NEC initially increases with \( m \), reaches a maximum, and subsequently decreases as \( m \) is increased further. Consequently, for an intermediate range of \( m \), the NEC remains non-negative over the entire radial domain for the chosen parameter values.
	
	The dependence on the normalization parameter \( \omega \) in Model~III also differs qualitatively from that observed in Models~I and II. In this case, the NEC decreases monotonically with increasing \( \omega \) across the spacetime. Nevertheless, a sufficiently large value of \( \omega \) is required to prevent NEC violation near the throat. From the numerical plots, the lowest value of \( \omega \) for which the NEC remains non-negative in the throat region is approximately \( \omega \simeq 0.7 \).
	
	\section{Conclusion}
    
    In this work, we have investigated three configurations within the framework of $f(R,\Box R)$ gravity, namely the standard Morris--Thorne wormhole metric, its deformative extension, and a further generalization incorporating a time-dependent scale factor. From the analysis of the energy condition profiles, it is observed that, prior to the introduction of the time evolution scale factor, the null energy condition (NEC) is asymptotically violated, although there exist narrow regions in the vicinity of the wormhole throat where the NEC is satisfied.

    With the inclusion of the time-dependent scale factor, the behavior of the NEC undergoes a notable change. The corresponding profiles indicate that, for most values of the parameters $m$ and $\omega$, the NEC remains satisfied throughout the spacetime, with only certain parameter choices leading to deviations from this behavior. 

    Furthermore, an additional feature emerges in the time-evolving case: during the late-time phase (i.e., for higher values of $t$), the NEC approaches zero while remaining positive, whereas in the early-time phase (lower values of $t$), it attains comparatively larger positive values. This trend is consistently observed across all considered functional forms of $f(R,\Box R)$.

    \bibliographystyle{unsrt}
    \bibliography{reference2}
    
\end{document}